\documentclass[12pt]{article}

\usepackage{latexsym}
\usepackage{graphicx}
\usepackage{amsmath}
\usepackage{amssymb}
\usepackage{mathrsfs}

\usepackage{epsfig}

\def\beq{\begin{equation}}
\def\eeq{\end{equation}}
\def\bea{\begin{eqnarray}}
\def\eea{\end{eqnarray}}

\begin{document}

\begin{titlepage}

\vspace*{1cm}
\begin{center}
{\bf \Large {A little quantum help for cosmic censorship \\ and a step beyond all that}}

\bigskip \bigskip \medskip

{\bf Nikolaos Pappas}

\bigskip
{\it Division of Theoretical Physics, Department of Physics,\\
University of Ioannina, Ioannina GR-451 10,  \\ Greece}

\bigskip\bigskip\bigskip

\begin{abstract}
The hypothesis of cosmic censorship (CCH) plays a crucial role in classical general relativity,
namely to ensure that naked singularities would never emerge, since it predicts that whenever a singularity
is formed an event horizon would always develop around it as well, to prevent the former from interacting
directly with the rest of the Universe. Should this not be so, naked singularities could eventually form,
in which case phenomena beyond our understanding and ability to predict could occur, since at the vicinity of
the singularity both predictability and determinism break down even at the classical (e.g. non-quantum) level.
More than $40$ years after it was proposed, the validity of the hypothesis
remains an open question. We reconsider CCH in both its weak and strong version, concerning point-like singularities,
with respect to the provisions of Heisenberg's uncertainty principle. We argue that the shielding of the
singularities from observers at infinity by an event horizon is also quantum mechanically favored, but ultimately
it seems more appropriate to accept that singularities never actually form in the usual sense, thus no naked
singularity danger exists in the first place.
\end{abstract}

\bigskip\bigskip\bigskip\bigskip\bigskip

\bigskip\bigskip\bigskip\bigskip\bigskip
\bigskip\bigskip\bigskip\bigskip\bigskip
\bigskip\bigskip\bigskip\bigskip\bigskip

Email: npappas@cc.uoi.gr

\end{center}

\end{titlepage}

\section{Introduction}
Singularities, conceived as space-time regions, where curvature (as described by scalar
invariant quantities like $R_{\mu\nu\rho\sigma}R^{\mu\nu\rho\sigma}$) blows up to exceed
any possible upper bound, are one of the most problematic notions in Physics. After all,
strictly speaking, if the space-time metric is ill-behaved at a certain point, then the latter
should not be considered as part of that space-time at the first place. Nevertheless, it is this metric that we rely on to try to describe
the properties of that point. We overcome this incoherence by considering an augmented space-time
that contains such singular points as ideal boundary points attached to the ordinary, well-behaved
manifold. Since the
space-time structure breaks down at singularities while, at the same time, physical laws presuppose space and time
to develop and manifest themselves, naked singularities would be sources of lawlessness, absurdity and uncontrollable
information, therefore an anathema for our perception of the Universe.
Even worse, Hawking and Penrose have shown that the emergence of
singularities is inevitable in a very large class of universe types, where sufficiently reasonable conditions are
satisfied\footnote{The theorem actually goes as: Let $\emph{M}$,$\emph{g}_{ab}$ be a time-oriented spacetime satisfying the conditions\\
$(1)$ $\emph{R}_{ab}\emph{V}^{a}\emph{V}^{b}\geq 0$ for any non-spacelike $\emph{V}^{a}$. \\
$(2)$ The timelike and null generic conditions are fulfilled. \\
$(3)$ There is no closed timelike curve. \\
$(4)$ At least one of the following holds:\\
\hspace*{1cm} $(4a)$ There exists a compact achronal set without edge. \\
\hspace*{1cm} $(4b)$ There exists a trapped surface. \\
\hspace*{1cm} $(4c)$ There is a $\emph{p}\epsilon \emph{M}$ such that the expansion of the future directed null
geodesics through $\emph{p}$ \\
\hspace*{1cm}becomes negative along each geodesic.\\
Then $\emph{M}$,$\emph{g}_{ab}$ contains at least one incomplete timelike or null geodesic.}
\cite{HawkPen-singularities}.
Since all of them are redeemed in our Universe too, singularities are expected with certainty to form in the latter as well.
In order to deal with these ``monsters'' Penrose proposed the famous cosmic censorship hypothesis (CCH) \cite{Pen-CCH1}. The weak version of the hypothesis (w-CCH)
suggests that observers at infinity can never directly see a singularity the latter being at all times clothed by an absolute event horizon,
whereas its strong version (s-CCH) states that an observer cannot have any direct interaction with a singularity at any time or place \cite{Pen-CCH2}.
Because of cosmic censorship then, a naked singularity should never occur except, conceivably, for some special
configurations, which are not expected to occur in an actual astrophysical circumstance
(for a thorough and enlightening presentation of the issues concerning singularities and CCH see \cite{Geroch-1968a}, \cite{book-Oxford}, \cite{book-Chicago}).

It should be noted here that CCH does not stem from some well established physical law or mathematical theorem. Rather, it is a
convenient hypothesis, that considering the catastrophic impact of the alternative, we gladly accept as (probably)
true. Soon after it was proposed, it was declared as one of the most important open questions in classical general relativity \cite{Pen-CCH2},
whose derivation remains obscure until nowadays (see \cite{review1} \cite{review2} for reviews on the work done sofar).
Initially, arguments supporting the idea were based largely on geometry and issues concerning causality,
usually expressed in terms of TIFs and TIPs (terminal indecomposable futures/pasts respectively) \cite{TIF & PIF} (see again \cite{book-Chicago}). Penrose was
able to derive inequalities involving black hole masses and horizon radii \cite{inequalities1} in support of his hypothesis, which interestingly
enough were shown to hold true in a series of different situations \cite{inequalities2}. Moreover, CCH was proven to be valid in various specific space-times
\cite{limited CCH}.
At the same time
sceptics were trying to construct counterexamples in which naked singularities could emerge \cite{counterexamples},
however the majority of those examples presupposed very special and idealized conditions to hold (thus least possible to occur in a realistic Universe) so the credibility of CCH was far from being fatally undermined by them.
Soon it was evident that a quantum treatment was necessary. Besides, the problematic way we describe singularities
represents much more our lack of understanding about their true nature, namely the laws of quantum gravity that
presumably takes over when radii of space-time curvature of the order of Planck length are attained, than their actual
behavior. It was proposed (and hoped) by many scientists that the inclusion of quantum phenomena in the picture of
gravitational collapse would be the answer to all our difficulties to cope with singularities. In fact, quantum mechanics
has been proven very successful in resolving many of the counterexample gedankenexperiments in favor of CCH. More specifically it was used to
show that it is impossible to over-spin or over-charge a maximal Kerr black hole to produce a naked singularity (a procedure first considered in \cite{Wald})
\cite{QM-vs-counterexamples}.

Following this line of thinking we try here to engage quantum mechanics in the treatment of point-like singularities
lying at a finite distance (as opposed to singularities lying at infinity or thunderbolts). The key idea
proposed is to make appeal to Heisenberg's uncertainty principle,
\begin{eqnarray}
\Delta x \cdot \Delta p \geq 1 \hspace{2em} (\rm in \hspace{0.5em} natural \hspace{0.5em} units \hspace{0.5em} where \hspace{0.5em} G = c = \hbar = 1) \label{uncertainty}
\end{eqnarray}
which we consider as the most fundamental feature of
every natural system, and check the constraints it imposes, concerning the properties of systems that involve singularities.

\section{Weak censorship revisited}
Point-like singularities are expected to form because of the unstoppable collapse of matter that occurs when too large a mass
is concentrated in too small a volume. The volume of these singularities would effectively tend to zero by definition, thus they should
occupy a single point of space-time. In the case of a naked singularity, an observer at infinity (that is at sufficiently large distance away from it
so as to be in an asymptotically flat region of space-time) would in principle be able to determine its position with
arbitrarily high accuracy by e.g. direct observation.
When we make a measurement with uncertainty $\Delta x \rightarrow 0$ concerning the position of a quantum system, however, the
uncertainty principle states that we have to end up with
complete lack of knowledge about its momentum (that is $\Delta p \rightarrow \infty$), therefore about its energy as well.
We argue, though, that this is not the case with naked singularities. Since in principle they can be of arbitrarily large mass, one reasonably expects
that the actual procedure of determining their position could not change their momentum significantly. Furthermore, even though quantum gravity is
necessary to describe the singularity \emph{per se}, it is legitimate to anticipate that general relativity is sufficiently accurate to
describe space-time at macroscopic distances away from it. Then, it would be possible to ``weigh'' the singularity
by observing potential gravitational lensing effects (see for example \cite{Kwirtata} for details) or through measuring the trajectory, speed and acceleration of test bodies that get
attracted by it etc. This way the mass/energy of the singularity would be known with uncertainty at most of the order of the mass itself ($\Delta M \sim M$).
All these mean that the existence of a naked singularity, apart from all other undesired consequences,
would also violate the uncertainty principle. The conundrum gets settled
when the provisions of the w-CCH is taken into account. The existence of an event horizon of radius $r_h \sim M$, that emerges because
of the warping of the space-time continuum by the singularity mass itself and thus exists in every kind of black hole type,
means that the actual position of the singularity can be determined with uncertainty at least $\Delta x \sim r_h$.
Then we get from eq. (\ref{uncertainty}) that $\Delta p \gtrsim \frac{1}{M}$ and consequently we find for the singularity energy
the inequality $E \gtrsim \frac{1}{M^3}$,
which obviously is perfectly compatible with measuring its mass/energy with $\Delta M \sim M$. In this sense w-CCH not only is necessary
to make general relativity self-consistent, but has a strong quantum support as well.

\section{Strong censorship revisited}
What about s-CCH then? It is not hard to imagine a situation where a very large and massive system is in question
(e.g. the central region of a galaxy), a trapped surface has already formed while observers living on a planet within
the trapped region exist and expect quantum mechanics to hold at all times until they crash into the singularity,
that will develop some time in their future.
Even though the soon-to-form singularity would remain unseen by observers at infinity
(so w-CCH is satisfied) an observer inside the horizon would actually encounter a naked singularity (being at the same time at
a significantly large distance away from it). All arguments presented in the paragraphs
above hold true for this observer too, so a paradox rises. The s-CCH is established to resolve the paradox by predicting that an observer
would never actual see the singularity, but since it doesn't
provide us with a mechanism capable of deterring this interaction, looks more like the expression of a hope than a
constraint imposed by some physical law.
The only way out, then, is to admit that the notion of unstoppable collapse
is wrong and, consequently, no point-like singularity is formed at all. Quantum effects should get so enhanced, at Planck scales, that
would manage to counterbalance the gravitational contracting forces to stop the collapse and prevent singularities from forming in the way we consider
them to do today \footnote{For example the confinement of matter in an ever decreasing volume, which means that it would acquire an ever
increasing momentum/energy, according, once again to the uncertainty principle, so that it would end up behaving like a highly
energetic gas whose pressure would constantly grow to counterbalance eventually the contraction is a plausible mechanism to be explored in a work to come.}.

This approach, namely the expectation that no singularity forms eventually, finds good support from a very interesting result by
Geroch which crudely goes as follows: when a manifold admits a Cauchy surface (as it is the case for the majority of physically reasonable
space-times), then it also admits a global time function $t$, that increases along every future-oriented timelike curve, which can be
chosen so that every $t = const.$ surface is a Cauchy one. However, Cauchy surfaces cannot intersect the singularity and thus there is
no time at which the singularity exists \cite{Geroch-1970b}.

To sum up, revisiting CCH on the grounds of the uncertainty principle, we end up that w-CCH should hold true. However, since by itself,
it is insufficient to make the overall picture self-consistent, it is needed that s-CCH also applies. Yet the latter in its turn imposes so strict
restrictions, that, as a way out, one quite naturally arrives to admit that singularities never emerge in the usual sense,
rendering CCH, in all its versions, unnecessary in the first place.

\end{document}